\documentclass[12pt]{article}
\usepackage{amssymb}
\usepackage{graphics}
\usepackage{epsfig}

\DeclareFontFamily{U}{rsf}{} \DeclareFontShape{U}{rsf}{m}{n}{
  <5> <6> rsfs5 <7> <8> <9> rsfs7 <10-> rsfs10}{}
\DeclareMathAlphabet\Scr{U}{rsf}{m}{n}

\catcode`\@=11
\def\@citex[#1]#2{%
\if@filesw \immediate \write \@auxout {\string \citation {#2}}\fi
\@tempcntb\m@ne \let\@h@ld\relax \def\@citea{}%
\@cite{%
  \@for \@citeb:=#2\do {%
    \@ifundefined {b@\@citeb}%
      {\@h@ld\@citea\@tempcntb\m@ne{\bf ?}%
      \@warning {Citation `\@citeb ' on page \thepage \space undefined}}%
      {\@tempcnta\@tempcntb \advance\@tempcnta\@ne%
      \@tempcntb\number\csname b@\@citeb \endcsname \relax%
      \ifnum\@tempcnta=\@tempcntb %Number follows previous--hold on to it
        \ifx\@h@ld\relax%
          \edef \@h@ld{\@citea\csname b@\@citeb\endcsname}%
        \else%
          \edef\@h@ld{\ifmmode{-}\else--\fi\csname b@\@citeb\endcsname}%
        \fi%
      \else%   %  non-successor--dump what's held and do this one
        \@h@ld\@citea\csname b@\@citeb \endcsname%
        \let\@h@ld\relax%
      \fi}%
    \def\@citea{,\penalty\@highpenalty\,}%
  }\@h@ld
}{#1}}

\def\@citeb#1#2{{[#1]\if@tempswa , #2\fi}}
\def\@citeu#1#2{{$^{#1}$\if@tempswa , #2\fi }}
\def\@citep#1#2{{#1\if@tempswa , #2\fi}}

\def\bcites{         % cite with []'s
        \catcode`\@=11
        \let\@cite=\@citeb
        \catcode`\@=12
}

\def\upcites{         % cite with exponents
        \catcode`\@=11
        \let\@cite=\@citeu
        \catcode`\@=12
}

\def\plaincites{      % cite without brackets
        \catcode`\@=11
        \let\@cite=\@citep
        \catcode`\@=12
}

\newcount\hour
\newcount\minute
\newtoks\amorpm
\hour=\time\divide\hour by 60 \minute=\time{\multiply\hour by 60
\global\advance\minute by-\hour}
\edef\standardtime{{\ifnum\hour<12 \global\amorpm={am}%
        \else\global\amorpm={pm}\advance\hour by-12 \fi
        \ifnum\hour=0 \hour=12 \fi
        \number\hour:\ifnum\minute<10 0\fi\number\minute\the\amorpm}}
\edef\militarytime{\number\hour:\ifnum\minute<10
0\fi\number\minute}

\def\draftlabel#1{{\@bsphack\if@filesw {\let\thepage\relax
   \xdef\@gtempa{\write\@auxout{\string
      \newlabel{#1}{{\@currentlabel}{\thepage}}}}}\@gtempa
   \if@nobreak \ifvmode\nobreak\fi\fi\fi\@esphack}
        \gdef\@eqnlabel{#1}}
\def\@eqnlabel{}
\def\@vacuum{}
\def\marginnote#1{}
\def\draftmarginnote#1{\marginpar{\raggedright\scriptsize\tt#1}}
\def\draft{
        \pagestyle{plain}
        \overfullrule=2pt
        \oddsidemargin -.5truein
        \def\@oddhead{\sl \phantom{\today\quad\militarytime} \hfil
        \smash{\Large\sl DRAFT} \hfil \today\quad\militarytime}
        \let\@evenhead\@oddhead
        \let\label=\draftlabel
        \let\marginnote=\draftmarginnote
        \def\ps@empty{\let\@mkboth\@gobbletwo
        \def\@oddfoot{\hfil \smash{\Large\sl DRAFT} \hfil}
        \let\@evenfoot\@oddhead}
        \def\@eqnnum{(\theequation)\rlap{\kern\marginparsep\tt\@eqnlabel}%
        \global\let\@eqnlabel\@vacuum}  }

\def\section{\@startsection {section}{1}{\z@}{3.ex plus 1ex minus
 .2ex}{2.ex plus .2ex}{\large\bf}}
\def\subsection{\@startsection{subsection}{2}{\z@}{2.75ex plus 1ex minus
 .2ex}{1.5ex plus .2ex}{\bf}}

\def\appendix{{\newpage\section*{Appendix}}\let\appendix\section%
        {\setcounter{section}{0}
        \gdef\thesection{\Alph{section}}}\section}

\def\abstract{\if@twocolumn
\section*{Abstract}
\else %\small
\begin{center}
{\bf Abstract\vspace{-.5em}\vspace{0pt}}
\end{center}
\quotation \fi}

\catcode`\@=12

\newcommand{\beq}{\begin{equation}}
\newcommand{\eeq}{\end{equation}}
\newcommand{\beqa}{\begin{eqnarray}}
\newcommand{\eeqa}{\end{eqnarray}}

\newcommand{\be}{\begin{eqnarray}}
\newcommand{\ee}{\end{eqnarray}}
\newcommand{\nn}{\nonumber}

\def\lae{\mathrel{\mathop{\smash{\lower .5 ex \hbox{$\stackrel<\sim$}}}}}
\def\lae{\mathrel{\mathop{\smash{\lower .5 ex \hbox{$\stackrel>\sim$}}}}}

\def\Tr{{\rm Tr}}

\def\l:{\mathopen{:}\,}
\def\r:{\,\mathclose{:}}

\catcode`\@=12

\bcites

\catcode`\@=11
\def\theequation{\arabic{equation}}
\@addtoreset{equation}{section} \@addtoreset{footnote}{section}
\@addtoreset{footnote}{subsection} \catcode`\@=12

\newcommand{\eqn}[1]{(\ref{#1})}
\def\Dslash{\,\,{\raise.15ex\hbox{/}\mkern-12mu D}}
\def\Dbarslash{\,\,{\raise.15ex\hbox{/}\mkern-12mu {\bar D}}}
\def\delslash{\,\,{\raise.15ex\hbox{/}\mkern-9mu \partial}}
\def\delbarslash{\,\,{\raise.15ex\hbox{/}\mkern-9mu {\bar\partial}}}
\def\pslash{\,\,{\raise.15ex\hbox{/}\mkern-9mu p}}
\def\calDslash{\,\,{\raise.15ex\hbox{/}\mkern-12mu {\cal D}}}

\begin{document}
\pagestyle{plain} \setcounter{page}{1}
\newcounter{bean}
\baselineskip16pt

\begin{titlepage}

\begin{center}

\hfill\today\\

\vskip 3.0 cm {\Large The Quantum Dynamics of Heterotic Vortex
Strings} \vskip 1 cm
{\large David Tong}\\
\vskip 1cm
{\sl Department of Applied Mathematics and Theoretical Physics, \\
University of Cambridge} \\
{\tt d.tong@damtp.cam.ac.uk}

\end{center}

\vskip 1.5 cm
\begin{abstract}
We study the quantum dynamics of vortex strings in ${\cal N}=1$
SQCD with $U(N_c)$ gauge group and $N_f=N_c$ quarks. The classical
worldsheet of the string has ${\cal N}=(0,2)$ supersymmetry, but
this is broken by quantum effects. We show how the pattern of
supersymmetry breaking and restoration on the worldsheet captures
the quantum dynamics of the underlying 4d theory. We also find
qualitative matching of the meson spectrum in 4d and the spectrum
on the worldsheet.
\end{abstract}
\end{titlepage}

%\tableofcontents
\setcounter{section}{1}
\section*{Introduction}

Vortex strings have proven to be a useful probe of
four-dimensional quantum dynamics. By studying the $d=1+1$ theory
on the string worldsheet, one may extract information about the
strongly coupled phase of the underlying four-dimensional theory.
For vortex strings in ${\cal N}=2$ supersymmetric theories, this
information includes the exact BPS mass spectrum, the
Seiberg-Witten curve, and the dimensions of chiral primary
operators at superconformal points \cite{sy,vstring,scvs,adam}.
For earlier, related work see \cite{nick,dht}; for further work in
this direction see
\cite{vib,auzzi,memono,other,others,bolog,syss,gsy,tasi,koji}.

The purpose of this short letter is to study the quantum dynamics
of vortex strings in ${\cal N}=1$ supersymmetric gauge theories.
The classical dynamics of these vortices was recently determined
in \cite{het}. The strings have ${\cal N}=(0,2)$ supersymmetry on
their worldsheet and were dubbed ``heterotic vortex strings". In
this paper we focus on $U(N_c)$ SQCD with $N_f=N_c$ fundamental
flavors. The low-energy quantum physics of this theory  was
understood some years ago by Seiberg in terms a quantum
deformation of the moduli space of vacua \cite{seiberg}. We will
show that this four-dimensional behavior can be reproduced by
studying the $(0,2)$ worldsheet dynamics of the vortex string. The
relevant physics on the worldsheet is dynamical supersymmetry
breaking and supersymmetry restoration. We will also find
qualitative matching of the meson spectrum in 4d and the spectrum
of the worldsheet.

Before jumping into the details, let us pause to review the basic
philosophy. When studying the quantum dynamics of solitons there
are typically two different scales in the game: $\Lambda$, the
scale of strongly coupled quantum effects and $M$, the scale of
the symmetry breaking that supports the soliton. The approach one
must take to study soliton dynamics depends on the ratio of these
two scales. When $\Lambda \gg M$, one must first deal with the
quantum dynamics in four dimensions and subsequently look for
solitons in the low-energy effective theory. In contrast, when $M
\gg \Lambda$, one should instead look for solitons in the
microscopic four-dimensional theory and then subject these
solutions to a traditional semi-classical quantization. Although
these approaches are valid in different regimes, comparing the
results can sometimes be profitable, especially if certain details
do not depend on the ratio $\Lambda/M$. For example, the
equivalence of Liouville theory and the two-dimensional black hole
\cite{hk} was understood in this way through the study of domain
walls \cite{mmotw}. In the following pages we will examine the
quantum dynamics of vortex strings using these two different
approaches. The punchline is that the two routes commute, and the
semi-classical quantization of the $(0,2)$ worldsheet theory
mirrors the quantum dynamics of the four-dimensional theory.
Stated another way, one can recover the 4d quantum dynamics  by
studying the 2d dynamics of the string.

\section*{SQCD: $N_f=N_c$}

Our focus in this paper is ${\cal N}=1$ SQCD, with $SU(N_c)$
gauge group and $N_f=N_c\equiv N$ fundamental flavors. Each flavor
consists of two chiral superfields, $Q$ and $\tilde{Q}$,
transforming in the ${\bf N}$ and $\bar{\bf N}$ representations of
the gauge group respectively. The classical theory has global
symmetry group
\be SU(N)_L\times SU(N)_R\times U(1)_A\times U(1)_B\times
U(1)_R\label{4dsym}\ee
under which the chiral superfields $Q$ transform as $({\bf N},{\bf
1}, 1,1/N,0)$ while $\tilde{Q}$ transform as $({\bf 1},\bar{\bf
N}, 1,-1/N,0)$. Note that the charge assignment refers to the
scalar component of $Q$ and $\tilde{Q}$; the fermionic components
do transform under the $U(1)_R$ R-symmetry, as does the gaugino.
The R-symmetry is non-anomalous. However $U(1)_A$ does suffer from
an anomaly, leaving a remnant ${\bf Z}_{2N}$.

The low-energy physics of this theory is well understood
\cite{seiberg}. The classical theory has a moduli space of vacua,
given by solutions to the D-flatness conditions. These solutions
are parameterized in a gauge invariant fashion by the meson chiral
superfield,
\be M_{ij}=\tilde{Q}_i Q_j\ee
together with a pair of baryon chiral superfields,
\be B=\epsilon_{a_1\ldots a_{N_c}}Q^{a_1}_1\ldots
Q^{a_{N_c}}_{N_c}\ \ \ ,\ \ \ \tilde{B}=\epsilon_{a_1\ldots
a_{N_c}}\tilde{Q}^{a_1}_1\ldots \tilde{Q}^{a_{N_c}}_{N_c}\ee
These are not independent. They obey the classical constraint
\be \det M - B\tilde{B}=0\label{clascons}\ee
The light fields in the classical theory are $M$, $B$ and
$\tilde{B}$ subject to the constraint \eqn{clascons}. The
resulting manifold has singularities at $\tilde{B}=B=0$ and ${\rm
rank}(M)<N-2$. These singularities reflect the existence of new
massless gluons which emerge when the symmetry breaking is less
than maximal.

The situation in the quantum theory is different. The classical
constraint \eqn{clascons} is corrected \cite{seiberg},
\be \det M - B\tilde{B}=\Lambda_{4d}^{2N}\label{qcons}\ee
where $\Lambda_{4d}^{2N} = \mu^{2N} e^{-4\pi^2/e^2(\mu)+i\theta}$
is proportional to the one-instanton action. The manifold defined
by \eqn{qcons} is smooth. The singularities of the classical
moduli space have been resolved, reflecting the confining nature
of the quantum theory

\subsection*{\it Vortices in the Low-Energy Theory}

Let us now describe how vortices appear in the low-energy theory.
The $SU(N_c)$ theory does not have the topology to support vortex
strings. To introduce vortices we deform the theory by gauging the
$U(1)_B$ baryon symmetry. Of the low-energy fields, $M$ is neutral
under $U(1)_B$ while $B$ and $\tilde{B}$ have charge $+1$ and $-1$
respectively. We introduce a Fayet-Iliopoulos (FI) parameter
$v^2>0$ for $U(1)_B$ which imposes the D-flatness condition on the
scalar fields,
\be |B|^2-|\tilde{B}|^2=v^2\label{d}\ee
Since $v^2>0$, we necessarily have $B\neq 0$ in vacuum. The
$U(1)_B$ invariant combination  $\tilde{B}B$ is then determined by
the meson vev through the constraint \eqn{clascons} or
\eqn{qcons}.

 The presence of the FI parameter ensures that the
$U(1)_B$ gauge symmetry is broken which, in turn, guarantees the
existence of vortex strings. One may still trust the low-energy
effective theory \eqn{qcons} provided one hits the 4d strong
coupling scale $\Lambda_{4d}$ before the Higgs mechanism induced
by $v^2$ kicks in. This means $\Lambda_{4d}\gg ev$.

The question that will concern us in this paper is: When are the
vortex strings BPS? The equations describing a BPS string are the
first order vortex equations,
\be F_{12}^{B} = e^2 (|B|^2-|\tilde{B}|^2-v^2) \ \ ,\ \ {\cal
D}_1B=i{\cal D}_2B\ \ ,\ \  {\cal D}_1\tilde{B}=i{\cal
D}_2\tilde{B}\ \ \ \label{bog}\ee
The key observation, which permeates this paper, is that these
equations have solutions only when $\tilde{B}=0$
\cite{penin,ddt,davis}. This fact follows from a standard theorem
in mathematics which states that there exists no non-zero
holomorphic section of negative degree --- see, for example,
\cite{phases}. To repeat: BPS vortices exist only if
$\tilde{B}=0$. When $\tilde{B}\neq 0$, any vortex solution must
necessarily be non-BPS.

Let us examine the conditions for BPS vortices in both the
classical and quantum theories. In the classical theory, the
constraint \eqn{clascons} allows us to happily sit in the vacuum
$|B|^2=v^2$, with $\tilde{B}=M=0$. This vacuum is identified by
the surviving $SU(N)_L\times SU(N)_R$ global symmetry. It enjoys
BPS vortices. Indeed, BPS vortices continue to survive in any
vacuum with ${\rm rank}(M)<N$. However, once ${\rm rank}(M)=N$,
the constraint \eqn{clascons} requires $\tilde{B}\neq 0$ and the
vortex is no longer BPS.

Things are rather different once we take into account the 4d
quantum dynamics. In the vacuum $M=0$, with $SU(N)_L\times
SU(N)_R$ global symmetry, we must necessarily have $\tilde{B}\neq
0$ and the vortices break supersymmetry. To find a BPS vortex, we
must now sit in a vacuum with $\det M = \Lambda^{2N}$. We learn
that, as far as the vortex strings are concerned, the 4d quantum
dynamics both giveth and taketh away. Quantum effects induce both
supersymmetry breaking, in which classically BPS vortices no
longer preserve supersymmetry, as well as supersymmetry
restoration, in which classically non-BPS vortices are rendered
BPS in the full quantum theory.

\section*{The Microscopic Vortex Theory}

We now return to the original $SU(N)$ gauge theory. We again gauge
$U(1)_B$ and introduce a FI parameter $v^2$. This induces vevs for
the squarks that break the $U(N)$ gauge group completely. This
time we work in the regime $ev \gg \Lambda_{4d}$, ensuring that
gauge group is spontaneously broken before strong coupling effects
take hold. In this regime, we may treat the vortex string in the
$U(N)$ theory in a semi-classical manner. We will find that by
studying dynamical supersymmetry breaking on the string worldsheet
we will be able to reproduce the quantum deformation \eqn{qcons}.

We start by discussing the string dynamics in the classical vacuum
$|B|^2=v^2$ and $M=\tilde{B}=0$. In this vacuum the classical
vortex is BPS, preserving one half of the four supercharges.  The
microscopic theory of the vortex string in this vacuum was
determined in \cite{het}. It is given by the ${\cal N}=(0,2)$
supersymmetric ${\bf C}\times {\bf CP}^{N-1}$ sigma-model. The
factor ${\bf C}$ describes the translational and Goldstino modes
of the string. The ${\bf CP}^{N-1}$ part describes the internal
orientation modes of the vortex in color and flavor space
\cite{vib,auzzi}. The theory is most simply described in terms of
a gauged linear sigma-model for $N$ homogeneous coordinates
$\phi_i$
\be {\cal L}_{\rm vortex} &=& |{\cal D}_m\phi_i|^2 +
2i\left(\bar{\xi}_{-i}{\cal D}_+\xi_{-i} + \bar{\xi}_{+i}{\cal
D}_-\xi_{+i}\right) +\frac{\theta}{2\pi}\epsilon_{\mu\nu}\partial^\mu A^\nu \nn\\
&&\ \ \ \ - \ D(|\phi_i|^2-r) -(\bar{\zeta}_-\bar{\xi}_{+i}\phi_i +
\bar{\phi}_i\xi_{+i}{\zeta}_-)\ \ \ \ \label{lag} \ee
The Lagrange multiplier $D$ imposes the constraint $\sum_i |\phi_i|^2 =r$.
Dividing by the $U(1)$ gauge action
$\phi_i\rightarrow e^{i\alpha}\phi_i$ reduces the theory to a sigma-model
with ${\bf CP}^{N-1}$ target space of size \cite{vib,others}
\be r = \frac{2\pi}{e^2}\label{r}\ee
The $\theta$-angle on the worldsheet is inherited from the
$\theta$-angle in the four-dimensional bulk \cite{vstring,others}.
Note that the Grassmannian Lagrange multiplier $\zeta_-$ restricts
the right-moving fermions to lie in the tangent space:
$\phi_i\bar{\xi}_{+i}=0$. In contrast the left-moving fermions
$\xi_{-i}$ are unconstrained\footnote{There is a subtlety here.
The left-moving fermi zero modes $\xi_{-i}$ are non-normalizable
due to an infra-red divergence. Nonetheless, they are needed to
avoid a sigma-model anomaly on the worldsheet. These problems can
be circumvented by considering the four-dimensional theory a torus
of area $A \gg 1/e^2v^2, 1/\Lambda^{2}$.}. The $(0,2)$ ${\bf
CP}^{N-1}$ sigma-model can be thought of as a deformation of the
more familiar ${\cal N}=(2,2)$ ${\bf CP}^{N-1}$ sigma-model
\cite{witquark,ddds,phases} in which the auxiliary
``$\sigma$-field" is removed, together with an auxiliary fermion.
In particular, the $(0,2)$ ${\bf CP}^{N-1}$ sigma-model \eqn{lag}
has no four-fermi interaction.

The worldsheet Lagrangian \eqn{lag} is invariant under the global symmetry group,
\be SU(N)_L\times SU(N)_R\times U(1)_A\times U(1)_R\ee
inherited from the subgroup of the 4d symmetry \eqn{4dsym} preserved in the
vacuum. (The worldsheet
$SU(N)_L$ actually descends from the unbroken diagonal combination
of the 4d $SU(N)_L$ and the gauge group $SU(N)$). The $\phi_i$
transform as $({\bf N}, {\bf 1},+1,0)$; the right-handed fermions
$\xi_{+i}$ transform as $({\bf N},{\bf 1}, +1,-1)$, while the
left-handed fermions $\xi_{-i}$ transform as $({\bf 1},{\bf N},
-1,-1)$. The $U(1)_R$ symmetry is non-anomalous, while the
$U(1)_A$ symmetry suffers an anomaly, leaving a ${\bf Z}_{2N}$
remnant. This mimics the behavior of the 4d symmetry. The agreement between
the anomaly structure in 4d and on the worldsheet is no coincidence. In
both cases, the $U(1)_A$ anomaly can be understood by examining the fermi zero
modes carried by the relevant instanton. Yet, the instanton in the 2d and 4d
theories are different
descriptions of the same object: the worldsheet instanton can be thought of as the
Yang-Mills instanton, trapped to lie within the vortex string
\cite{vstring}. As one sends $v^2\rightarrow 0$, the vortex string
melts away but the worldsheet instanton remains, changing smoothly
into the Yang-Mills instanton. The number of fermi zero modes is dictated
by an index and remains unchanged in this limit, ensuring that the anomalies
in the bulk and on the worldsheet agree.

\subsubsection*{\it Dynamical Supersymmetry Breaking}

We now turn to the quantum dynamics of the microscopic worldsheet
theory. We will show that the Lagrangian \eqn{lag} dynamically
generates an expectation value for the Lagrange mutliplier $D$.
Since $D$ is the auxiliary field in an ${\cal N}=(0,2)$ vector
multiplet, and may be written as a supersymmetry variation $\delta
\zeta_-= \epsilon_- D$, this means the theory dynamically breaks
supersymmetry. In fact, the expectation value for $D$ was already
shown 30 years as a simple application of the $1/N$ expansion in
the bosonic sigma-model \cite{daddabose,witquark}, and it is not
hard to check that the presence of fermions in the $(0,2)$ model
do not change this conclusion\footnote{In contrast, supersymmetry
is not broken in the ${\cal N}=(2,2)$ supersymmetric model. This
can be traced to the presence of an extra bosonic field $\sigma$
which bears the burden of the expectation value instead of $D$.}.
Ignoring the fermions, one first integrates out the bosonic fields
$\phi_i$ to leave the partition function
\be Z_{\rm bose}=\int dD\,dA_\mu\ \exp\left(-N\,\Tr\log
\left[-(\partial_\mu+iA_\mu)^2-D\right]+i\int d^2x\ Dr
+\frac{\theta}{2\pi}\epsilon_{\mu\nu}\partial^\mu
A^\nu\right)\nn\ee
The Lorentz invariant ground state has $A_\mu=0$, with $D$ sitting at the
stationary point,
\be ir + N\int
\frac{d^2k}{(2\pi)^2}\,\frac{1}{k^2-D+i\epsilon}=0\ee
The integral can be performed exactly to give the supersymmetry breaking
expectation value,
\be D = 4 \Lambda^2_{2d}\label{sgone}\ee
where the strong coupling scale on the worldsheet is given at
one-loop by $\Lambda_{2d}^N= \mu^N e^{-2\pi r+i\theta}$. Recalling
the relationship \eqn{r}, we can match the $\Lambda$ parameters on
the worldsheet and the bulk by first running $e^2$ down to the
scale of the vortex tension, and subsequently running $r$ down to
strong coupling: $\Lambda_{4d}^{2N} = v^N \Lambda_{2d}^N$. The
dynamical supersymmetry breaking on the worldsheet mirrors the
behavior we saw in studying the four-dimensional quantum dynamics:
the classical vortex is BPS, but is rendered non-BPS by quantum
effects. We will shortly see what becomes of the vortex in the
other 4d vacua.

\subsubsection*{\it The Spectrum}

One of the most striking feature of vortex strings in ${\cal N}=2$
theories is that the BPS mass spectrum of the worldsheet coincides
with the BPS spectrum of the bulk theory in the phase with $v^2=0$
\cite{nick,dht,sy,vstring}. What is the story in our ${\cal N}=1$
theory? There are no longer BPS particle states with their
associated non-renormalization theorems, and exact agreement of
the spectrum would be optimistic. Nonetheless, one might hope to
match qualitative features of the spectrum such as quantum
numbers. Most notably, one would hope to see the confining nature
of the ${\cal N}=1$ theory from the perspective of the worldsheet.

Why do we expect even qualitative agreement between the spectrum
of the worldsheet and the spectrum of the 4d theory in the
confining phase? Let us return momentarily to the situation with
$ev \ll \Lambda_{4d}$. At distance scales $\ll 1/ev$, the 4d
particle spectrum is largely unaffected by the Higgs mechanism.
Some of these particles are likely to form weakly bound states
with the vortex string. As we now increase $ev/\Lambda_{4d}$,
those bound states which remain light ($\ll ev$) will  appear as
internal excitations of the microscopic worldsheet theory.

The spectrum of the 4d theory contains both baryons and mesons.
Once we turn on $v^2\neq 0$, the baryons are all screened in the
bulk and unlikely to form bound states with the vortex. This
leaves us with the spectrum of mesons, consisting of the massless
fields $M_{ij}=\tilde{Q}_iQ_j$ transforming in the bi-fundamental
representation of the $SU(N)_L\times SU(N)_R$ global symmetry
group, together with massive fields $Q^\dagger_iQ_j$ and
$\tilde{Q}_i\tilde{Q}^\dagger_j$ transforming in the singlet and
adjoint representations. Because of confinement, there are no
physical states transforming in the fundamental representations of
the global symmetry group. How much of this is seen by the string?

Let us firstly recall some basic facts about ${\bf CP}^{N-1}$
sigma-models. The purely bosonic ${\bf CP}^{N-1}$ model is well
known to exhibit confinement \cite{daddabose,witquark}. This was
previously noted in the context of the vortex theory in
\cite{others}. Confinement occurs because the auxiliary $U(1)$
gauge field in \eqn{lag} becoming dynamical after integrating out
$\phi_i$ and the Coulomb potential grows linearly in
two-dimensions. This means that the bosonic model contains no
physical states transforming in the ${\bf N}$ or $\bar{\bf N}$
representations of the global $SU(N)$ symmetry group. All physical
states live in either singlet or adjoint representations.

Typically the physics of the ${\bf CP}^{N-1}$ model changes
drastically in the presence of fermions and often the model no
longer confines \cite{witquark}. The mechanism by which this
happens is rather interesting. As we have seen above, if we couple
$N$ fermions, each charged under the worldsheet $U(1)$ gauge
field, then the $U(1)_A$ axial symmetry is anomalous, broken to
the discrete subgroup ${\bf Z}_{2N}$. Moreover, under the right
conditions --- a caveat that will be clarified below --- strong
coupling effects induce a condensate of fermion bilinears
$\bar{\xi}_i\xi_i= \bar{\xi}_{+i}\xi_{-i}-\bar{\xi}_{-i}\xi_{+i}$,
\be \langle \bar{\xi}_{i}\xi_{i}\rangle \sim \Lambda_{2d}
\label{fcond}\ee
This condensate spontaneously breaks the anomaly free discrete
symmetry ${\bf Z}_{2N}\rightarrow {\bf Z}_2$. This ensures the
existence of $N$ isolated vacua of the model, together with the
associated particle states arising from kinks which interpolate
between the different vacua. These kinks transform in the
fundamental representation of $SU(N)$ and, in fact, can be
identified with the original degrees of freedom $\phi_i$
\cite{witquark}. The condensate \eqn{fcond} occurs in the ${\cal
N}=(2,2)$ ${\bf CP}^{N-1}$ model where it is a key element in
matching the spectrum of the vortex worldsheet to the spectrum of
${\cal N}=2$ four-dimensional theories. In that case, neither the
worldsheet theory, nor the 4d theory, confines.

What of our ${\cal N}=(0,2)$ ${\bf CP}^{N-1}$ model? The
worldsheet theory includes fermions, so we may expect the
mechanism above to again be at play. However, the four-dimensional
physics is certainly confining. What's going on? In fact, the
condensate \eqn{fcond} cannot form in the $(0,2)$ ${\bf CP}^{N-1}$
sigma-model. The reason is that, in contrast to the $(2,2)$ model,
the theory has a chiral $SU(N)_L\times SU(N)_R$ flavor symmetry
which would be broken by \eqn{fcond}.  However, there can be no
spontaneous symmetry breaking in two dimensions: the ground state
wavefunction spreads to preserve all global symmetries in the
vacuum \cite{mw,cole}. The preservation of the continuous
$SU(N)_L\times SU(N)_R$ acts as a guardian of the discrete
$Z_{2N}$ symmetry and neither are broken in the vacuum. This
ensures that the worldsheet theory confines, and all particle
states transform in the adjoint, singlet or bi-fundamental
representations of $SU(N)_L\times SU(N)_R$, in broad agreement
with the 4d meson spectrum.

More is known about the worldsheet theory. It can be shown that at
at least $N^2-1$ of the states are massless. This follows by
treating the large $N$ theory as QED${}_2$ with $N$ massless
flavors \cite{bs,halp,cris} to which is almost reduces. (The
presence of the constraint for the right-moving fermions means it
differs slightly, but this does not change the conclusions here).
It also follows on general grounds from the $SU(N)_L\times
SU(N)_R$ chiral symmetry \cite{nice}. These massless modes are not
Goldstone bosons but rather are related to the fact that the
theory lies in the Berezinski-Kosterlitz-Thouless phase, with
power-law decay of the correlator \cite{wthir,lnrev}
\be \langle \bar{\xi}_{i}(x)\xi_{i}(x)\
\bar{\xi}_{i}(0)\xi_{i}(0)\rangle \sim x^{-2/N-2 }\ee
It is natural to identify these massless modes as threshold bound
states of the chiral mesons $M_{ij}$ with the vortex strings. The
remaining particles on the worldsheet have mass of order
$\Lambda_{2d}$ and presumably originate from bound states of the
massive 4d mesons.

\subsection*{\it Exploring the Higgs Branch}\label{higgsbranch}

We have studied the vortex dynamics in the vacuum $M=0$, where
quantum effects dynamically break supersymmetry on the vortex
worldsheet. Let us now place ourselves in other vacua of the 4d
theory. We first need to understand how the classical vortex
dynamics responds to changing the meson vev $M$. The answer to
this question in ${\cal N}=2$ theories was given in \cite{scvs}
and a similar analysis applies here. The key physics involved is
similar to that described above. Turning on an expectation value
for $M$ in the classical 4d theory necessarily means turning on a
vev for $\tilde{Q}$.  But, by the same argument given after
equation \eqn{bog}, there can be no solution to the non-Abelian
first order vortex equations if $\tilde{Q}$ is sourced and carries
a non-trivial profile. In the non-Abelian theory, the vortex can
avoid this by sitting in part of the gauge group unaffected by the
vev for $\tilde{Q}$. This removes some of the moduli $\phi_i$. It
is simple to adapt the argument given in \cite{scvs} to the
present case. One finds that BPS vortices are given by the zeroes
of the worldsheet potential,
\be \delta{\cal L}_{\rm vortex} =
\bar{\phi}_i\,\frac{M^\dagger_{ij}M_{jk}}{v^2}\,\phi_k +
{\bar{\xi}}_{i-}\frac{M_{ij}}{v}\,\xi_{j+} + {\rm
h.c.}\label{hb}\ee
The overall constant in front of this potential is undetermined,
while the factors of $v$ here are required on dimensional grounds.
It is simple to show that this deformation preserves ${\cal
N}=(0,2)$ supersymmetry. (For example, it may be written in
superspace language as a $(0,2)$ superpotential). There exist
zeroes of the potential in \eqn{hb}, subject to the D-term
constraint $|\phi_i|^2=r$  only when ${\rm rank}(M)< N$. This is
indeed the condition for the existence of classical BPS vortices.

Let us now see how this conclusion is changed by quantum dynamics
on the worldsheet. We again work in the $1/N$ expansion and
integrate out $\phi_i$. As before, the smoking gun for broken
supersymmetry is the expectation value of $D$. Including now the
extra term \eqn{hb}, we find that equation \eqn{sgone} is replaced
by,
\be \det\left(\frac{M^\dagger M}{v^2}+D\,1_N\right)=
4^N\Lambda_{2d}^{2N}\ee
To leading order in $1/N$, $D$ has vanishing expectation value,
and supersymmetry is restored, providing that we tune the the
meson expectation value to
\be \det M \sim v^N\Lambda_{2d}^{N} \sim \Lambda_{4d}^{N} \ee
where $\sim$ refers to the fact that we have not fixed the overall
constant. This requirement for BPS vortices reproduces  the
analysis of vortices in the quantum deformed moduli space
\eqn{qcons}. It is noteworthy that when ${\rm rank}(M)=N$, the
classical worldsheet theory spontaneously breaks supersymmetry,
yet quantum effects may restore supersymmetry in the infra-red:
the classical non-BPS vortex is rendered BPS.

\section*{Soft Breaking from ${\cal N}=2$}

There is another limit in which we can profitably understand the
dynamical supersymmetry breaking on the worldsheet of the vortex
string, and that is in softly broken ${\cal N}=2$ theory. We
introduce a new scale $\mu$, which gives a mass
to the adjoint scalar $\Phi$ in the ${\cal N}=2$ vector multiplet. The
4d superpotential is given by
\be {\cal W}_{4d} = \sqrt{2}\sum_{i=1}^{N_f}\tilde{Q}_i(\Phi-m_i)Q_i +
\mu \Phi^2\ee
where we have also introduced complex masses $m_i$ for each flavor.

The 4d theory becomes strongly coupled at the scale $\Lambda_{N=2}
=\mu e^{-4\pi^2/e^2N+i\theta/N}$ which, one can check, is equal to
$\Lambda_{2d}$. When $\Lambda_{N=2}\gg ev$, we should first study
the strongly coupled dynamics in four-dimensions. Our goal is to
understand the expectation value of $\tilde{B}$ as a function of
$\Lambda_{N=2}$ and the masses $m_i$, for this will tell us when
BPS vortices exist.  We work in several steps. Let us first
consider the ${\cal N}=2$ theory with $\mu=0$. The theory has a
Coulomb branch, parameterized by the vev of $\Phi$. However, since
we ultimately wish to gauge $U(1)_B$ and turn on a FI parameter
$v^2$, we are not interested in the theory at an arbitrary point
on the Coulomb branch but rather in the special vacuum which lies
at the root of the baryonic Higgs branch. Classically this is
given by $\Phi={\rm diag}(m_1,\ldots,m_N)$ where $N$ quarks are
massless and may condense once $v^2\neq 0$. Quantum mechanically,
the information about the dynamics is encoded in  the
Seiberg-Witten curve, a rational curve in ${\bf C}^2$ whose
periods determine the couplings of low-energy fields. At the root
of the baryonic Higgs branch, this curve degenerates, reflecting
the presence of the massless quarks. It is given by,
\be F(v,t) = [\Lambda_{N=2}^Nt-\prod_{i=1}^N(v-m_i)^2][t-1]\equiv C_L(t,v)C_R(t) = 0
\label{swcurve}\ee
The next step is to introduce $\mu$, softly breaking the theory to
${\cal N}=1$. The low-energy dynamics is now captured  by a curve
in ${\bf C}^3$ instead of ${\bf C}^2$. (This is particularly
apparent in the five-brane picture where the curve is the brane,
and soft-breaking corresponds to rotating the brane in transverse
directions). For the degenerate curve \eqn{swcurve}, the effect of
the soft breaking is simple: the curve once again factorizes and
is given by \cite{ho,deboeroz}
\be C_{L}: &\ & \Lambda^N t = \prod_{i=1}^N(v-m_i)\ \ ,\ \ \omega=0 \nn\\
C_R: && t = 1\ \ ,\ \ \omega = 2\mu v
\ee
In \cite{ho,deboeroz}, the expectation of the baryonic operator $\tilde{B}B$ was
argued to be the distance between these two branches at $\omega=0$. This gives
us
\be \tilde{B}B = \Lambda^N - \prod_{i=1}^N(-m_i)\ee
The final step is to gauge $U(1)_B$ and turn on the FI parameter
$v^2$. Once again, the vortices are only BPS if $\tilde{B}=0$. The
condition for the existence of BPS vortices is therefore
\be  \Lambda^N = \prod_{i=1}^N(-m_i)\label{tune}\ee
Let us now see how to reproduce this result from the quantum
dynamics of the worldsheet. When $\Lambda_{N=2}\ll ev$, we should
treat the string semi-classically. In the limit $\mu=0$, where we
have 4d ${\cal N}=2$ supersymmetry, the low-energy dynamics of the
string is governed by the ${\cal N}=(2,2)$ ${\bf C}\times {\bf
CP}^{N-1}$ sigma-model \cite{vib,auzzi}, with a potential induced
by the masses $m_i$ for the 4d hypermultiplets
\cite{memono,sy,vstring}. The worldsheet action \eqn{lag} for the
${\bf CP}^N$ part is to be augmented with a scalar $\sigma$ and a
fermion $\zeta_+$ and the additional terms
\be \delta{\cal L}_{(2,2)\ {\rm vortex}} = \sum_{i=1}^N\
|\sigma-m_i|^2|\phi_i|^2 - \bar{\zeta}_+\phi_i\bar{\xi}_{-i} +
\bar{\xi}_{-i}(\sigma-m_i) \xi_{+i}+{\rm h.c.}\ee
As in  previous sections, we may integrating out the chiral
multiplets containing $\phi_i$. This now yields an effective
twisted superpotential for the twisted chiral multiplet $\Sigma$
whose lowest component is the scalar $\sigma$
\cite{ddds,phases,hh}
\be {\cal W}(\Sigma) = -\frac{1}{2\pi} \sum_{i=1}^N(\Sigma-m_i)
\log\left[\left(\frac{\Sigma - m_i}{M_{UV}}\right) - 1
\right]-(r+i\theta/2\pi)\Sigma\label{spot}\ee
with $M_{UV}$ the renormalization subtraction point. The scalar
potential energy is given by
$(K_{\sigma\bar{\sigma}})^{-1}|\,{\partial {\cal
W}}/{\partial\sigma}|^2$ where $K$ is the (unknown) K\"ahler
potential. Assuming $K$ is smooth, the $N$ vacuum states lie at
\be \frac{\partial {\cal W}}{\partial\sigma} = 0 \ \ \Rightarrow \
\ \prod_{i=1}^N(\sigma-m_i)=\Lambda_{2d}^N\label{spotvac}\ee
When the 4d theory is softly broken to ${\cal N}=1$ by the
addition of the mass $\mu$, the worldsheet theory is softly broken
to ${\cal N}=(0,2)$ in way that was determined in \cite{het}. When
$\mu \ll \Lambda_{2d}$, it is appropriate to view the breaking
within the effective worldsheet theory determined by \eqn{spot}.
The ${\cal N}=(0,2)$ deformation mixes a would-be Goldstino mode from
the ${\bf C}$ part of the ${\cal N}=(2,2)$ theory with the ${\bf CP}^{N-1}$
fields. The net effect on the worldsheet is to induce a mass for $\sigma$
\cite{het}, so that the scalar potential now reads
\be V = (K_{\sigma\bar{\sigma}})^{-1}\left|\frac{\partial {\cal
W}}{\partial\sigma}\right |^2 + |\,\mu\sigma|^2\ee
We once again see that the worldsheet dynamics mirrors the 4d
quantum dynamics. For generic values of the masses $m_i$, the
vacuum has non-vanishing energy and supersymmetry on the
worldsheet is spontaneously broken. However, when the masses are
tuned to satisfy \eqn{tune}, the solution to \eqn{spotvac} is
given by $\sigma=0$, and a vacuum with zero energy is recovered.
This is in full agreement with the analysis of baryon expectation
values in four-dimensions.

\section*{Acknowledgement}
My thanks to Nick Dorey and Matt Wingate
for several very useful discussions. I'm
supported by the Royal Society.

\end{document}